\documentclass[10pt]{article}
\usepackage{graphicx}

\def\Title#1{\begin{center} {\Large #1 } \end{center}}
\def\Author#1{\begin{center}{ \sc #1} \end{center}}
\def\Address#1{\begin{center}{ \it #1} \end{center}}

\newcommand\pubblock{\rightline{\begin{tabular}{l} Proceedings of the Fifth Annual LHCP\\ \pubnumber\\
         \pubdate  \end{tabular}}}

\newenvironment{Abstract}{\begin{quotation} \begin{center} 
             \large ABSTRACT \end{center}\bigskip 
      \begin{center}\begin{large}}{\end{large}\end{center} \end{quotation}}

\newenvironment{Presented}{\begin{quotation} \begin{center} 
             PRESENTED AT\end{center}\bigskip 
      \begin{center}\begin{large}}{\end{large}\end{center} \end{quotation}}

\def\Acknowledgements{\bigskip  \bigskip \begin{center} \begin{large}
             \bf ACKNOWLEDGEMENTS \end{large}\end{center}}

\def\beq{\begin{equation}}
\def\eeq#1{\label{#1}\end{equation}}
\def\eeqn{\end{equation}}

\def\beqa{\begin{eqnarray}}
\def\eeqa#1{\label{#1}\end{eqnarray}}
\def\eeqan{\end{eqnarray}}

\let\bar=\overbar

\def\Dslash{\not{\hbox{\kern-4pt $D$}}}
\def\dslash{\not{\hbox{\kern-2pt $\del$}}}

\def\msb{{\bar{\ssstyle M \kern -1pt S}}}

\usepackage{color}

\newcommand\pubnumber{ CMS-CR-2017/226 \\ FERMILAB-CONF-17-366-CMS }

\newcommand\pubdate{\today}

\def\affiliation{
Particle Physics Division \\
Fermi National Accelerator Laboratory\\
Batavia, Illinois, USA, 60510-0500\\
On behalf of the CMS Experiment}

\begin{document}

\large
\begin{titlepage}
\pubblock

\vfill
\Title{Highlights and Perspectives from the CMS Experiment }
\vfill

\Author{ Joel Nathan Butler }
\Address{\affiliation}
\vfill
\begin{Abstract}

\begin{flushleft}

In 2016, the Large Hadron Collider provided proton-proton collisions at 13 TeV center-of-mass energy and achieved very high luminosity and reliability.  The performance of the CMS  Experiment in this running period   and a selection of recent physics results are presented. These include precision measurements and searches for new particles. The status and prospects for data-taking in 2017 and a  brief  summary of the highlights of the High Luminosity (HL-LHC) upgrade of the CMS detector are also presented.

\end{flushleft}

\end{Abstract}
\vfill

\begin{Presented}
The Fifth Annual Conference\\
 on Large Hadron Collider Physics \\
Shanghai Jiao Tong University, Shanghai, China\\ 
May 15-20, 2017
\end{Presented}
\vfill
\end{titlepage}
\def\thefootnote{\fnsymbol{footnote}}
\setcounter{footnote}{0}
%

\normalsize 


\section{Introduction}

This report covers the performance of the Large Hadron Collider (LHC)  and Compact Muon Solenoid (CMS) Experiment\cite{CMSdet} in 2016; recent physics results from CMS including precision measurements of the Higgs boson, first observed  by the ATLAS and CMS collaborations in 2012\cite{Aad:2012tfa},\cite{Chatrchyan:2012ufa} and top quark properties, b-physics, and standard model (SM) phenomena; searches for new particles; the status and prospects for data-taking in 2017; and a  brief  summary of the highlights of the proposed High Luminosity (HL-LHC) upgrade of CMS. It will conclude with a summary and outlook.


In 2015, the Large Hadron Collider operated successfully for the first time at 13 TeV and with 25 ns bunch spacing. In 2016, the LHC concentrated on providing the experiments with very high integrated luminosity. This was achieved by the LHC's reaching a peak initial luminosity of 1.5$\times$10$^{34}$cm$^{-2}$s$^{-1}$  and providing very high availability for physics of $\sim$50\%.  

The peak luminosity, corresponding to a number of simultaneous interactions per beam crossing (pileup) of $\sim$40,  exceeded the design goal for the LHC and for which CMS was designed. In anticipation of a high peak and integrated luminosity in 2016 and beyond, CMS made various improvements and upgrades to the detector. These included improved Level 1 Muon Trigger and Calorimeter Triggers. 


In 2015, the cryogenic system, or ``cold-box" of the CMS Solenoid did not perform well, leading to significant periods in which the magnet could not be operated. A very aggressive program of cleaning the system of impurities, carried out in early 2016, was successful and the CMS Solenoid and cryogenics had almost no down time in 2016. 

With these improvements, CMS was able to fully exploit the high integrated luminosity provided by the LHC. 
A total of  40.2$fb^{-1}$  of luminosity was delivered to CMS. This  exceeded the goal for the run of 25 $fb^{-1}$. During this run, each CMS sub-detector had at least 96\% of its channels operational.  For the whole run, 92.5\% of the luminosity delivered by the LHC was recorded and 95\% of the recorded data,  35.9$fb^{-1}$, was certified as suitable for all physics analyses.

In order to be able to record data efficiently with the high luminosities of 2016,  CMS  implemented several innovative  analysis techniques. Only two will be described briefly, Particle Flow (PF) and Pile UP per Proton Interaction (PUPPI). 

\begin{figure}[htb]
\centering
\includegraphics[height=2in]{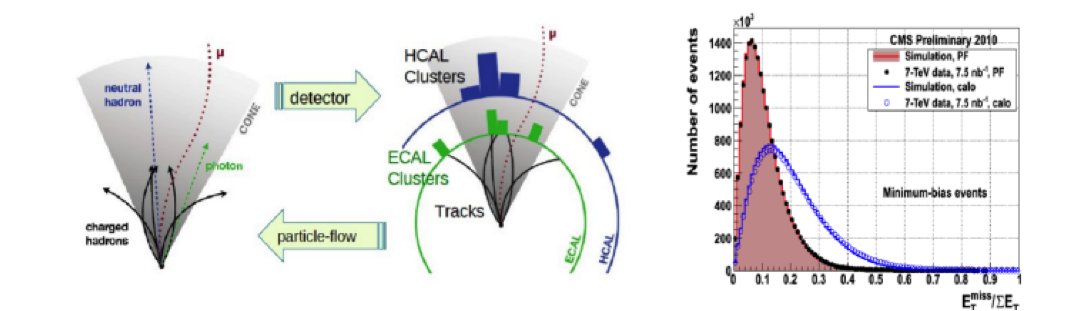}
\caption{ The middle figure shows how the particles within a cone appear to the detector and  the leftmost figure gives  its interpretation as physics objects by the PF algorithm. The righthand picture compares the missing $E_{T }$ in minimum bias events, which is typically small, obtained by PF and conventional 
``calorimeter"-only-based (CALO) methods. The PF distribution is narrower and more peaked towards zero
\cite{PF}.}
\label{fig:PF}
\end{figure}

\begin{figure}
\centering
\begin{minipage}{0.4\textwidth}
\includegraphics[width=0.8\textwidth]{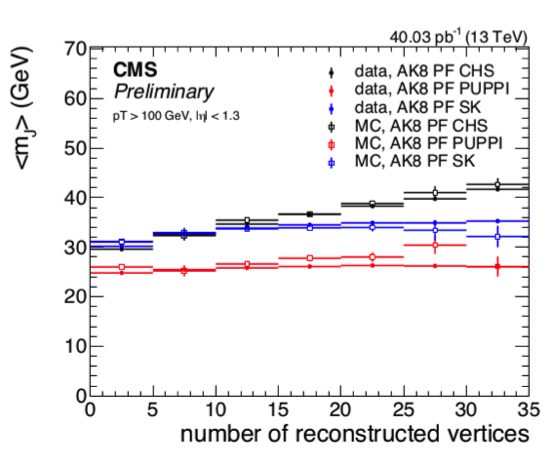}
\end{minipage}
\hfill
\begin{minipage}{0.4\textwidth}
\includegraphics[width=0.8\textwidth]{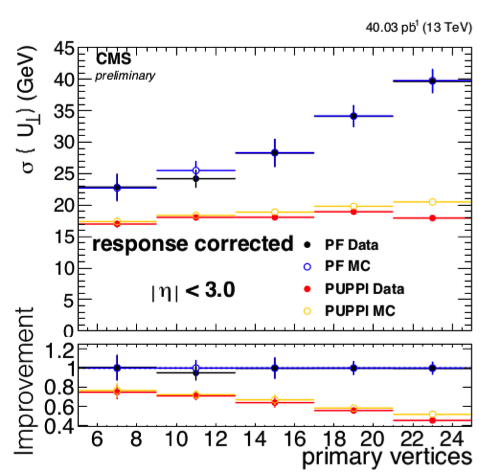}
\end{minipage}
\caption{ Average jet  mass, $M_{j}$ (left)and  transverse energy resolution of a jet recoiling against a Z boson (right) plotted vs the pileup. PUPPI is less sensitive to pileup than other techniques\cite{PUPPI}}.
\label{fig:PUPPI}
\end{figure}

In PF\cite{PF}, information from all detectors is used optimally to reconstruct each primary physics object. For example, in PF,  tracking is used together with calorimetry to reconstruct showers, jets, etc, unlike in more traditional approaches where only calorimetry is used.This is shown schematically in 
Figure~ \ref{fig:PF}.  PF works well in CMS because particles are well-separated in the large tracker volume and the 3.8T magnetic field; CMS has an all-silicon tracker extending from  4cm to 1m in radius, giving excellent track resolution and coverage with the  ablity to reconstruct particles with momenta as low as a few hundred MeV/c. The electromagnetic calorimeter has excellent resolution and high granularity.  Although the hadron calorimeter does not have high granularity,   only 10\% of the total energy in a jet comes from neutral hadrons that are measured only by   the hadron calorimeter. The charged hadron energy can be determined from the momentum measurement of the tracks. Tracks which never reach the calorimeter but are reconstructed in the tracker are included in jets. PF  produces a big improvement in jet energy resolution, $\tau$ identification,  and helps untangle the high $P_{T}$ interaction of interest from tracks and energy from pileup interactions. The basic idea of the PUPPI\cite{PUPPI} algorithm is to weight particle momenta according to the probability of not having pileup activity around it. The performance of the algorithm is shown in Figure~\ref{fig:PUPPI}, which demonstrates that the mass and $E_{T}$ resolution of jets from the primary interaction are independent of the number of pileup interactions.

 

\section{Recent Physics Results}

CMS has continued its prolific production of physics results. As of  the beginning of LHCP2017, 605 papers have been submitted to peer reviewed journals. Thirty-nine new results were released for the Moriond conferences this spring. Another 18 results were released before this conference. 
Many results are based on the full 35.9$fb^{-1}$ obtained in 2016. Here, we present a few highlights of recent results, many of which were shown  during this conference.

\subsection{Precision Measurements}

Because of its high luminosity and energy, the LHC produces massive numbers  of SM particles, whose properties are now being studied with unprecedented precision. The goal is to uncover subtle variations from SM behavior, which could signify new particles and forces in the quantum loops that contribute to higher order corrections.  This search for indirect evidence for new phenomena is often referred to as ``intensity frontier" physics in distinction to ``energy frontier" physics. The LHC, in addition to being the world's energy frontier collider, is also a superb intensity frontier machine for the top quark, b quark, W and Z bosons, and now  the Higgs boson. Here, we present some recent results on precision physics.

\subsubsection{Top Quark Properties}

Top quarks are produced in $t\bar{t}$ pairs by the strong interaction and singly by the electroweak interaction. The cross section for top pair production at the LHC at 13 TeV is more than 100 times the cross section at the Fermilab Tevatron at 1.96 TeV.  In fact, $t-\bar{t}$ rates at the LHC, requiring the production of more than 350 GeV/c$^{2}$ of mass, are quite comparable to the rates for $b-\bar{b}$ production at the so-called $B$-factories, as shown in Table~\ref{tab:factory}\cite{KEKb}. With the top pair rate exceeding  10 Hz, it is possible to address such issues as single and double differential cross sections,
rare (Flavor Changing Neutral Current) decays, CP violation, the top quark width, and more complex methods for measuring the mass, which may eventually produce better results than we have at present.
Top pair production  at 13 TeV CM energy is mainly (80\%) produced by gluons, providing important information on the  gluon distribution at relatively high $x_{F}$, up to  $\sim$0.25.

\begin{table}[h]
\begin{center}
\begin{tabular}{ccccc}  
Factory &  Quark &Cross section &  Luminosity & Rate\\
             &             & (nb)	        & (cm$^{-2}$s$^{-1}$)  &  (Hz)\\ \hline  
$B$ (KEKb)  &   Bottom     &     1.15 ($\Upsilon$(4S))     &     2.11 $\times$ 10$^{34}$ & 24.3\\
LHC &  Top    &       0.82 (inclusive $t\bar{t}$) &   1.5  $\times$ 10$^{34}$  & 12.3\\ \hline
\end{tabular}
\caption{ Comparison of $b-\bar{b}$ rates  at KEKb  and $t-\bar{t}$ rates at the LHC.}
\label{tab:factory}
\end{center}
\end{table}

\begin{figure}[htb]
\centering
\begin{minipage}{0.4\textwidth}
\includegraphics[width=\textwidth]{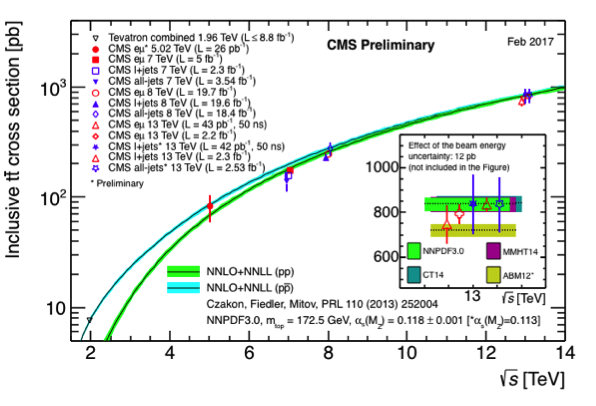}
\caption{ Top pair production total cross section vs. center of mass energy\cite{Top_xsec}. }
\label{fig:top_xsec}
\end{minipage}
\hfill
\begin{minipage}{0.4\textwidth}
\includegraphics[width=0.8\textwidth]{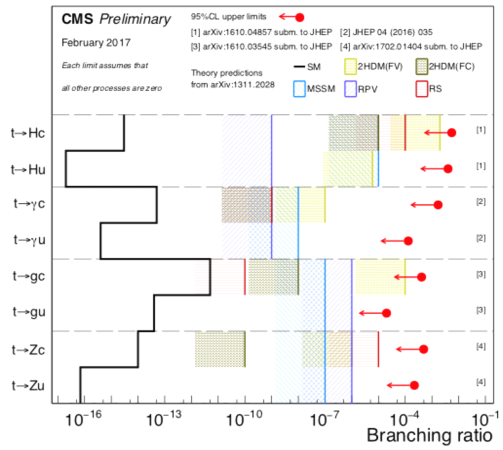}
\caption{ Limits on rare decays of the top quark \cite{Top_fcnc}.}
\label{fig:top_rare}
\end{minipage}
\end{figure}

The cross section for $t-\bar{t}$ production as a function of center of mass energy is shown in Fig.~\ref{fig:top_xsec}. The result for 13 TeV from CMS\cite{Top_xsec} is 833 $\pm$ 33 $pb$, in good agreement with theory. Single and double differential cross sections for key variables of top pair production are now also measured\cite{Top_1dif}, \cite{Top_2dif}. 

Already in 2011 and 2012 (Run 1), with the LHC running at 7 and 8 TeV in the center of mass, respectively, the first attempts to measure CP violation\cite{Top_CP}  and to search for rare decays of the top quark were undertaken. Fig.~\ref{fig:top_rare} shows recently published results on rare flavor-changing neutral current decays of top\cite{Top_fcnc}. While measurements at the LHC will not reach the level predicted by the SM, for some decays we will soon approach the predictions of various models of new physics  and will be able to confirm them or rule them out.  

\begin{figure}[htb]
\centering
\begin{minipage}{0.2\textwidth}
\includegraphics[width=\textwidth]{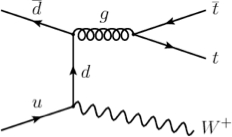}
\end{minipage}
\hfill
\begin{minipage}{0.2\textwidth}
\includegraphics[width=\textwidth]{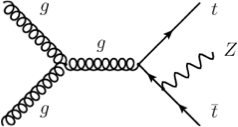}
\end{minipage}
\hfill
\begin{minipage}{0.4\textwidth}
\includegraphics[width=\textwidth]{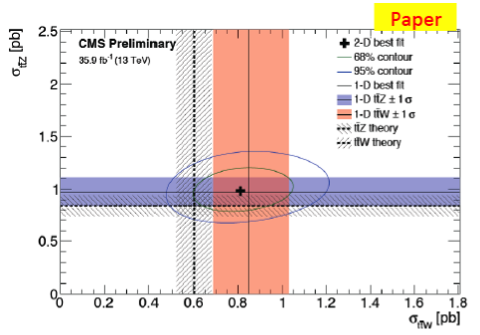}
\end{minipage}
\caption{ Feynman diagram for  producing  $ttW$ seen via same sign di-leptons (left) and $ttZ$ seen via 3 and 4 leptons (middle), along with the best fit values for the cross sections of $ttW$ and $ttZ$ (right)\cite{Top_ttv} }
\label{fig:ttWZ}
\end{figure}

Top pair production with an additional Z or W boson has also been observed. The process gives the best measurement of the coupling of the top quark to the $Z$ and also provides a measurement of these important background  processes for many searches for new particles. Feynman diagrams for the processes detected through 2 same-sign leptons and 3 or 4 leptons are shown in Fig.~\ref{fig:ttWZ}.  $t\bar{t}W$ is observed\cite{Top_ttv} with a significance of 5.9 $\sigma$ (4.7 $\sigma$ expected). The measured $t\bar{t} W$ signal strength is 1.22$^{+0.08}_{-0.09} (\rm{stat}) ^{+0.11}_{-0.13} ( \rm{sys})$. A fit to the combined cross sections of the $t\bar{t} W$ and $t\bar{t} Z$ is shown in Fig.~\ref{fig:ttWZ}.

\subsubsection{Higgs Boson Properties}

\begin{figure}[htb]
\centering
\begin{minipage}{0.4\textwidth}
\includegraphics[width=\textwidth]{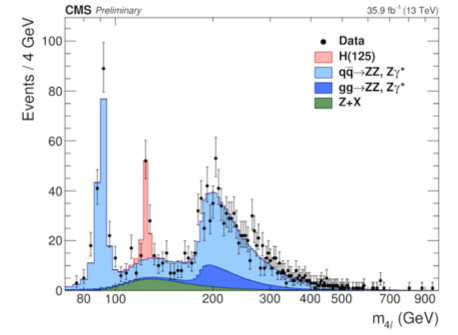}
\end{minipage}
\hfill
\begin{minipage}{0.4\textwidth}
\includegraphics[width=0.8\textwidth]{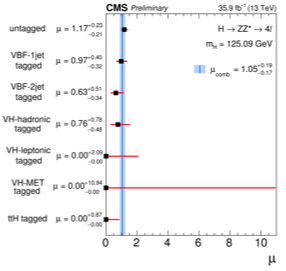}
\end{minipage}
\caption{  Mass spectrum of 4 muons from the 2016 data (left) and signal strength for the $HZZ^{*}$ channel broken down  by Higgs production mechanism (right)\cite{H_4L}.}
\label{fig:Higgs_4L}
\end{figure}

\begin{figure}[htb]
\centering
\begin{minipage}{0.25\textwidth}
\includegraphics[width=\textwidth]  {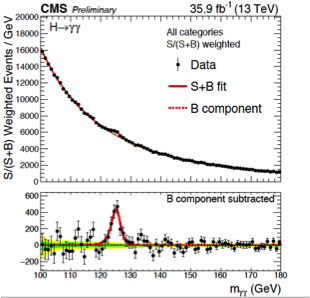}
\end{minipage}
\hfill
\begin{minipage}{0.30\textwidth}
\includegraphics[width=\textwidth] {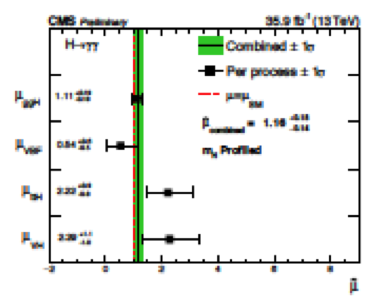}
\end{minipage}
\hfill
\begin{minipage}{0.20\textwidth}
\includegraphics[width=\textwidth] {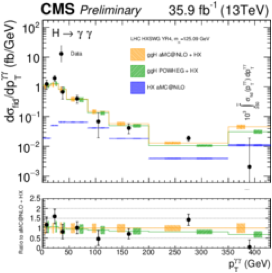}
\end{minipage}
\caption{ $\gamma \gamma$ mass spectrum (left); signal strength for various production modes (center); and $P_{T}$ distribution for Higgs production from di-photon decay (right) \cite{Higgs_gg}. }
\label{fig:Higgs_gamma}
\end{figure}

With more total luminosity and higher energy, larger samples of Higgs bosons are reconstructed, making more precise measurements and new studies possible. Fig.~\ref{fig:Higgs_4L}a\cite{H_4L} shows the mass spectrum of $ZZ^{*}\rightarrow 4 \  \rm{leptons}$. From this distribution the mass of the Higgs is measured to be 
$125.26\pm 0.20 (\rm{stat}) \pm 0.08 (\rm{syst})$,  a precision that exceeds the combined ATLAS and CMS result, $125.09\pm 0.21( \rm{stat}) \pm 0.1 (\rm{syst})$ using both $ZZ^{*}$ and $\gamma \gamma$ from Run 1\cite{Higgs_mass_r1}. The coupling strength of the Higgs to the $Z$ is broken down by production mechanism in ~Fig.~\ref{fig:Higgs_4L}b. The combined result is 1.05$^{+0.19}_{-0.17}$. Higgs decay to $\gamma \gamma$ is also studied\cite{Higgs_mass_r1}. Fig~\ref{fig:Higgs_gamma} shows the invariant mass distribution, the coupling strengths, and the differential cross section as a function of $P_{T}$\cite{Higgs_gg}.

The decay of the Higgs into $\tau^{+}\tau^{-}$ is reconstructed in four separate decay modes, $e\mu$,$e\tau_{h}$, $\mu\tau_{h}$, and $\tau_{h}\tau_{h}$ and in three production channels, 0-jet, Vector Boson Fusion (VBF), and ``boosted"\cite{Higgs_tautau}. Here  $\tau_{h}$ refers to specific targeted hadronic decay modes. The signal strength is 1.06$^{+0.25}_{-0.24}$  and the significance is 4.9$\sigma$, which, when combined with the result from CMS for Run 1, gives a significance that easily exceeds 5$\sigma$. This will constitute the first observation in a single experiment of the coupling of the Higgs to a fermion, to a lepton, and to a third generation family member. It also provides a validation of the Higgs Yukawa interaction. 

\begin{figure}[htb]
\centering
\begin{minipage}{0.4\textwidth}
\includegraphics[width=\textwidth]{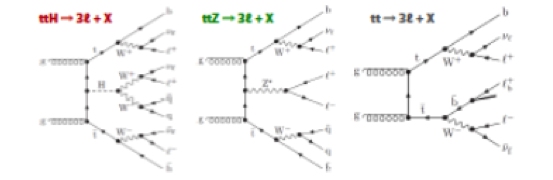}
\end{minipage}
\hfill
\begin{minipage}{0.4\textwidth}
\includegraphics[width=0.5\textwidth]{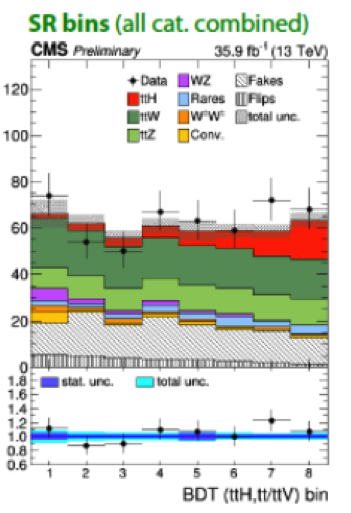}
\end{minipage}
\caption{ Feynman diagrams for multilepton signtures of ttH (left) and output of BDT showing excess of events from ttH (right) \cite{Higgs_tt}. }
\label{fig:figure15}
\end{figure}

CMS has also seen evidence for the $t\bar{t}$-Higgs coupling by studying multilepton final states\cite{Higgs_tt}. Feynman diagrams for some of the processes that contribute are shown in Fig.~\ref{fig:figure15}, which also shows the signal plot. The significance of the result in the 2016 data is 3.0$\sigma$ and when combined with Run 1 data is 3.3$\sigma$ (2.5 expected). The observed (expected) signal strength is 1.5$^{+0.5}_{-0.5}$ (1.0$^{+0.5}_{-0.4}$).

CMS also observed $WW$ scattering using two same-signed leptons and two jets. The observed(expected) significance is  5.9 (5.7) $\sigma$. The observed signal strength is 0.90$\pm$0.22\cite{WW_scat}. 

CMS searched as well for lepton flavour violating decays of the Higgs boson to $e\tau$ and $\mu\tau$  with the full 2016 statistics\cite{Higgs_rare_16}.  Previously, there was a  2.4$\sigma$ excess in $H\rightarrow \mu \tau$ in Run1 data \cite{Higgs_rare_15} but this was  not confirmed in the new data.

\subsubsection{P5$^{\prime}$ in $B^{0}\rightarrow K^{*}\mu^{+}\mu^{-}$}

\begin{figure}[htb]
\centering
\begin{minipage}{0.4\textwidth}
\includegraphics[width=\textwidth]{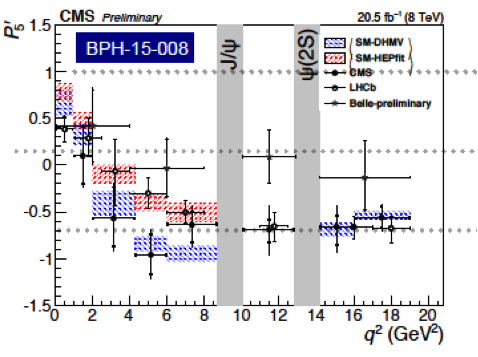}
\caption{ Angular coefficient P5$^{ \prime}$ vs. $q^{2}$, the dimuon invariant mass squared \cite{P5_prime}.  }
\label{fig:P5_prime}
\end{minipage}
\hfill
\begin{minipage}{0.4\textwidth}
\includegraphics[width=\textwidth]{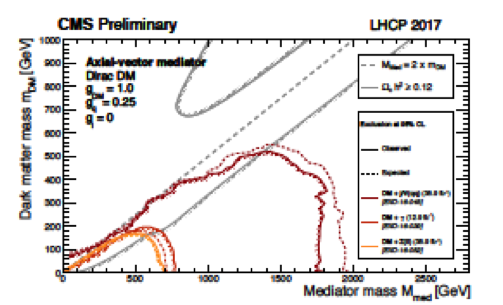}
\caption{ 95\% CL observed/expected exclusion regions  for MET-based DM searches in the lepto-phobic axial vector mode\cite{DM_search}.}
\label{fig:DM1}
\end{minipage}
\end{figure}

CMS measured  the P5$^\prime$ parameter and saw good agreement with the SM\cite{P5_prime}. This is an interference term in the angular distribution which is sensitive to contributions to BSM physics.The result is shown in Figure~\ref{fig:P5_prime}.

\subsection{Searches for New Particles}

{\bf Dark Matter:} Dark matter (DM) can be observed as missing $E_{T}$  in events with a large transverse momentum SM object. Searches for Mono-jets and mono-V(W,Z)  have been undertaken with the full 2016 dataset. Limits are based on Simplified Models of DM. An example of a recent result\cite{DM_search}  is shown in Fig.~\ref{fig:DM1}.


{\bf Supersymmetry:} Supersymmetry (SUSY) is an extension of the SM that resolves some of its open issues: the hierarchy problem, the unification of the couplings of strong and EWK forces, and the provision of a candidate for dark matter.  Many physicists expected  SUSY to be seen soon after LHC startup, with as little as 100$pb^{-1}$ of luminosity, and to be the first  major LHC discovery– before even the Higgs boson. 

Now with more than 40 $fb^{-1}$ of integrated luminosity at 13 TeV, SUSY has not yet been observed.  The reasons for this could be: the lightest SUSY particles  may be heavier than we thought; the pattern of decays may be more devious/obscure than we thought (coverage for 
R-Parity violating SUSY and long-lived particles are not as complete as for  MSSM or PMSSM searches); maybe SUSY does not cure all the above problems and there are several new physics scenarios, including SUSY, creating a complicated landscape in which we are currently  temporarily lost; or maybe SUSY is  just another promising idea that nature does not choose to follow.

CMS continues to pursue a broad program of SUSY searches. Nineteen searches have been completed  with the full 2016 CMS dataset, with several already submitted to journals. These 
probe different models;  inclusive signatures; strong and electroweak production; and 3rd generation sparticles (stops and sbottoms). Searches investigate different final states and  use several complementary analysis techniques.

\begin{figure}[htb]
\centering
\begin{minipage}{0.25\textwidth}
\includegraphics[width=\textwidth]  {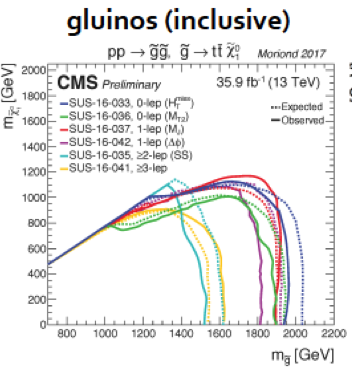}
\end{minipage}
\hfill
\begin{minipage}{0.30\textwidth}
\includegraphics[width=\textwidth] {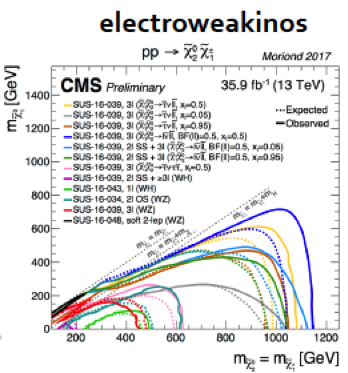}
\end{minipage}
\hfill
\begin{minipage}{0.20\textwidth}
\includegraphics[width=\textwidth] {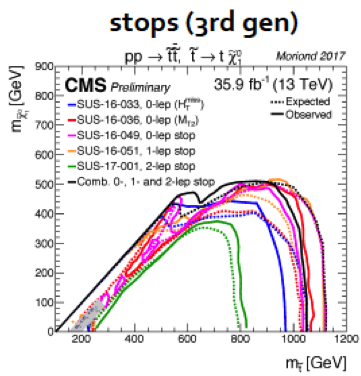}
\end{minipage}
\caption{ Limits on inclusive gluinos decaying into top and LSP (lleft); Electroweakino search (center); Limits for stop decaying into top and LSP (right) \cite{SUSY_gen}.}
\label{fig:figure13}
\end{figure}

Decay chains with two or four top quarks in the final state are promising signatures for SUSY\cite{SUSY_gen}. Of particular interest are  stop-quark decays to top and the Lightest Supersymmetric Partner (LSP). Stop was considered a very promising way of cancelling the contribution of the top quark loop to the Higgs mass. It is also possible for  gluinos to be produced and decay  through a top and a virtual stop going to top and LSP or  they can be produced and decay through a top and a real stop going to charm and LSP.
Special techniques are developed to identify  multiple hadronically-decaying top quarks over a large range of $P_{T}$. Both boosted and resolved jet reconstruction are used. Unfortunately, no signal is seen and stops are ruled out at 95\% CL up to masses of ~1 TeV for a neutralino/LSP less than about 430 GeV\cite{SUSY_stop}.


{\bf Vector-like quarks and top-partners:} Vector-like quarks have both left and right handed couplings to the $W$. Many models of BSM physics predict such particles. Since vector-like quarks could be quite massive, their SM decay products can receive high boosts. Special techniques are employed to include in the analyses two or more boosted jets which have merged to form ``fat" jets. CMS has recently done a search for the charge 2/3 vector-like top quark, T, produced in pairs. Specifically, the search is for
$T\bar{T} \rightarrow W^{+}b, W^{-}\bar{b} \rightarrow bl\nu \bar{b}q\bar{q}^{\prime}$. 
No signal was observed and a limit was established at 95\% confidence level of 1365 GeV (1245 GeV expected)\cite{VLQ_23}. A search was also undertaken for a charge 5/3 partner of the top quark decaying into $tW$, resulting in a limit of 1090 (990) GeV for right(left)-handed coupling\cite{VLQ_53}.


{\bf Long-lived particles:} Many BSM models predict long-lived particles. These result in distinctive properties such as displaced vertices, displaced conversions, displaced photons or jets, displaced leptons or di-leptons, heavy stable charged particles (HSCP) with very high ionization loss, or stopped HSCP.    Some of these can be observed by special searches, usually with special triggers. A recent example is a search by CMS for stopped long-lived particles. The signature is a high energy jet in the calorimeter that is out of time with collisions. This search excluded gluinos with lifetimes from 10$\mu$s to 1000s for gluino masses 
$ <$1379 GeV; and top squarks in this lifetime range with stop mass $<$ 740 GeV\cite{Long_lived}. This is an exciting new area of inquiry that has not been thoroughly explored at the LHC.

\section{Status of CMS for the 2017 run and beyond}

Run 2, which began in 2015, resumed in May of 2017 after an ``Extended Year-End Technical Stop" (EYETS) of $\sim$5 months ($\sim$2 more months than usual). Running at 13 TeV with luminosity as high as 1.9$\times$10$^{34}$ cm$^{-2}$s$^{-1}$ will produce roughly 
45-50 $fb^{-1}$ of integrated luminosity in each of 2017 and 2018. 

\begin{figure}[htb]
\centering
\begin{minipage}{0.4\textwidth}
\includegraphics[width=\textwidth]{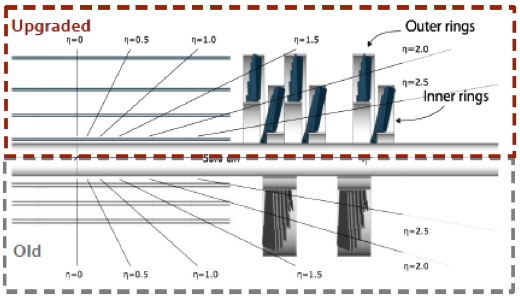}
\end{minipage}
\hfill
\begin{minipage}{0.4\textwidth}
\includegraphics[width=\textwidth]{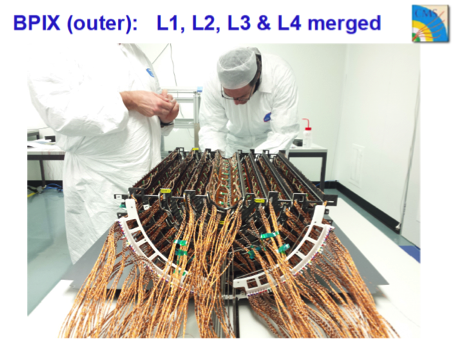}
\end{minipage}
\hfill
\caption{ Schematic of the Phase 1 upgrade detector layout compared to the original pixel detector (left) and photo of 1/2 of the new barrel pixel detector (4 layers) before integration into its supply tube (right).  }
\label{fig:pixel}
\end{figure}

In the EYETS, CMS installed a new pixel detector\cite{pixel_tdr} in anticipation of the rise in luminosity to  2$\times$10$^{34}$ cm$^{-2}$s$^{-1}$ with 25ns bunch crossing.  This corresponds to a pileup  of more than 50. The total rates and the pileup are twice what the original detector was designed to handle.  The new detector features more robust tracking: 4 layer coverage in the barrel and 3 disks in each endcap as compared to 3 layers and 2 disks that we had up to the  end of 2016;
faster readout able to run up to  2-2.5$\times$10$^{34}$ cm$^{-2}$s$^{-1}$ with almost no inefficiency (from hit loss) or dead time; and radiation hard enough to survive an integrated luminosity of 500 $fb^{-1}$. 
The new detector is much lighter, placing  much less material in front of the outer tracker and calorimetry. 
The inner barrel layer is closer to beam to provide  better primary and  secondary vertex resolution.
Figure~\ref{fig:pixel}a shows a schematic of the new detector and compares it with its predecessor.
Fig.~\ref{fig:pixel}b shows one half of the barrel pixel ready to be installed in its supply tube. At the time of LHCP, the detector had just been installed in CMS. In addition to the new pixel detector, improvements were made to the forward hadron calorimetry and the luminosity monitors.   A demonstrator for a Gaseous Electron Multiplier (GEM)  detector, a technology that will be used in future muon system upgrades, was also installed\cite{gem_tdr}.

The current upgrade of the CMS detector will continue in Long Shutdown 2 (LS2) in 2019 and 2020 with replacement of the photosensors, front-end, and back end electronics for the hadron calorimeter\cite{hcal_tdr}. It will then be able to operate efficiently with the luminosity planned for the next three years of LHC running. Around 2024, the LHC will shut down to install upgrades that allow it to reach a peak luminosity of at least 7.5$\times$10$^{34}$, corresponding to a pileup of more than 200 interactions per bunch crossing. We will run this way for about 10 years, accumulating an integrated luminosity of 3000-4500 $fb^{-1}$. This era of the LHC is referred to as the High Luminosity LHC (HL-LHC).

To handle such extreme pileup and radiation levels and to take advantage of the unprecedented sensitivity that will be possible,  much of the CMS detector has to be upgraded once again\cite{phase2_tp}. The goal of the HL-LHC upgrade is for the new CMS to be as efficient, and with as low background/fake-rates, at 200-250 pileup as we are today, and with extended acceptance in $\eta$. There will be a completely new all-silicon tracker with much higher granularity to preserve the reconstruction efficiency and keep the ``fake" rate low. It will be composed of more radiation-tolerant electronics and light-weight materials than the current tracking system. It will facilitate track reconstruction in the first level trigger, with a latency of about 4 $\mu$s. There will also be extended pixel detector coverage out to $\eta$= 3.8. The Barrel Electromagnetic  calorimeter will receive new electronics to handle the longer trigger latency and higher Level 1 output rate. To cope with  radiation damage, it will run at a lower temperature of -10$^{o}$C.
The Muon system electronics will be upgraded and new chambers will be added from $1.6  \,< \eta < \, 2.4$.
Muon tagging will be extended to  2.4$ < \eta <$ 3 with the addition of new chambers.
The endcap calorimeters, now consisting of an electromagnetic section employing $PbWO_{4}$ crystals and a plastic scintillator wavelength-shifting fibre hadron calorimeter, will be replaced with new high granularity Endcap Calorimeters whose electromagnetic section is made of plates of lead with read out by silicon detectors and a hadron calorimeter section composed of stainless steel plates read out also by planes of silicon detectors, followed by additional planes of plastic scintillators read out using Silicon Photomultpliers (SiPMs). This system provides a full energy measurement for electrons photons, charged hadrons, and jets, along with  precision measurement of the arrival time of the particles at the detector. With over 9 million readout channels and the ability to follow the full time and 3-D spatial development of showers, it is well-suited to handling high pileup in the forward direction.  

The two stage trigger scheme CMS has used up to now will be preserved but the first stage, Level 1 (L1) will have a latency of  12.5$\mu$s, enough to accommodate a track trigger,  and an output rate  to the next level of 750KHz. The next stage, the High Level Trigger (HLT), will process this data and output about 7.5 KHz of events to permanent storage for analysis. 

\begin{figure}[htb]
\centering
\begin{minipage}{0.4\textwidth}
\includegraphics[width=\textwidth]{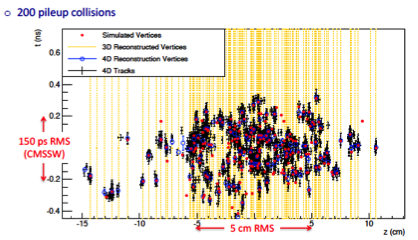}
\end{minipage}
\hfill
\begin{minipage}{0.4\textwidth}
\includegraphics[width=0.6\textwidth]{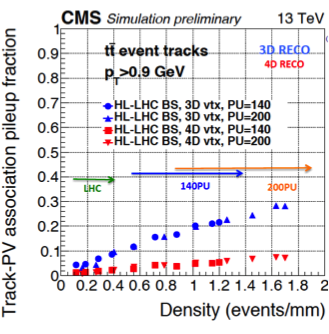}
\end{minipage}
\caption{ Collision time (y axis) vs interaction point in Z (x axis) (left)  and frequency of tracks falsely assigned to primary vertex vs pileup (density of interactions along beam axis)(right)\cite{mtd_ref}}
\label{fig:mtd}
\end{figure}

An additional detector will be added to ensure that CMS can cope with high pileup even if the LHC exceeds its design goals. This is the incorporation of a "MIP Precision Timing Detector"\cite{mtd_ref} in both the barrel and endcap regions. The barrel timing layer is located just outside the barrel tracker and the endcap timing layer is just in front of the endcap calorimeter. The detectors take advantage of the fact that the interactions in the same bunch crossing are spread out in time with an RMS of about 200 ps. The timing layers will have time  resolution of $\sim$ 30 ps, for $∣\eta∣< 3, \  p_{T} > 0.7 \  {\rm GeV}$.   By connecting tracks with vertices and the timing of showers, which is also approximately 30 ps, in the calorimeter, a factor of 4-5 in effective pile-up reduction can be achieved. Figure~\ref{fig:mtd}a shows a view of the vertices in a single crossing as they are distributed in time and in vertex Z along the beam. 
Figure~\ref{fig:mtd}b shows that without timing about 15\% of the vertices are merged but with timing this is reduced to 2.5\%. The discriminating power from timing is evident.

\section{Conclusions}

%

Now that the LHC is running at 13 TeV with high luminosity and availability,  our discovery potential is great.  
Discoveries may come in a  few months  or after several years. They might start with a  striking signal appearing in a single channel or they may appear as several  signals emerging slowly with initially low significance out of large backgrounds from a multiplicity of hiding places.  They may appear in scenarios we have long been exploring, e.g. SUSY or Extra Dimensions, or may surprise us with signatures that we are not even looking for today. As investigators/researchers into the unknown we need to step back and survey the big picture and look for new, untried  approaches or unexplored corners of our data. Today we have  $<$2\% of the ultimate LHC data in hand. It is our mission to explore and make discoveries  in this huge new expanse of scientific territory. Our future at CMS is bright and we are eagerly looking ahead to the challenges before us.

\Acknowledgements
The author wishes to thank the many members of the CMS Collaboration and their technical and administrative staffs whose efforts are summarized in this report. We congratulate our colleagues in the CERN accelerator departments for the excellent performance of the LHC. This document was prepared by the author using in part the resources of the Fermi National Accelerator Laboratory (Fermilab), a U.S. Department of Energy, Office of Science, HEP User Facility. Fermilab is managed by Fermi Research Alliance, LLC (FRA), acting under Contract No. DE-AC02-07CH11359.


\begin{thebibliography}{99}


\bibitem{CMSdet} CMS Collaboration, ``The CMS Experiment at the CERN LHC", {\it JINST} {\bf 3} (2008) S08004, doi:10.1088/1748-0221/3/08/S08004.

\bibitem{Aad:2012tfa} 
  G.~Aad {\it et al.}  [ATLAS Collaboration],
  Phys.\ Lett.\ B {\bf 716}, 1 (2012)
  [arXiv:1207.7214 [hep-ex]].
  
  
\bibitem{Chatrchyan:2012ufa} 
  S.~Chatrchyan {\it et al.}  [CMS Collaboration],
  Phys.\ Lett.\ B {\bf 716}, 30 (2012)
  [arXiv:1207.7235 [hep-ex]].


\bibitem{PF}  CMS Collaboration, ``Particle-flow reconstruction and global event description with the CMS detector", arXiv:1706.04965, CMS-PRF-14-001, CERN-EP-2017-110 ;

\bibitem{PUPPI} D. Bertolini, P. Harris, M. Low, and N. Tran, ``Pileup Per Particle Identification", 
{\it JHEP} {\bf 10} (2014) 059, doi:10.1007/JHEP10 (2014) 059, arXiv:1407.6013.

\bibitem{KEKb}  www.interactions.org/node/12633 and www.interactions.org/node/12633

\bibitem{Top_xsec} Top xsec LHC Top Physics Working Group, twiki.cern.ch/twiki/bin/view/LHCPhysics/LHCTopWGSummaryPlots

\bibitem{Top_1dif} CMS Collaboration, ``Measurement of differential cross sections for top quark pair production using the leptons+jets final state in proton-proton collisions at 13 TeV", Phys. Rev. D 95(2017) 092001, doi:10.1103/PhysRevD.95.092001, arXiv:1610.04191


\bibitem{Top_2dif} CMS Collaboration, ``Measurement of double differential cross sections for top quark pair production in pp collisions at $\sqrt{s} \ = \ 8 {\rm TeV}$ and impact on parton distribution functions",
Eur. Phys. J. C77 (2017) 459, doi: 10.1140/epjc/s10052-017-4984-5, arXiv:1703..01630


\bibitem{Top_CP} Top CP CMS Collaboration,``Search for CP violation in $t\bar{t}$ production and decay in proton-proton collisions at 8 TeV", doi:10.1007/JHEP03 (2017) 101,  arXiv:1611.08931

\bibitem{Top_fcnc} CMS Collaboration, Summary of FCNC upper limits at the 95\% CL from CMS searches at 8 TeV, twiki.cern.ch/twiki/bin/view/CMSPublic/PhysicsResultsTOPSummaryFigures

\bibitem{Top_ttv} CMS Collaboration, ``Measurement of top pair production in association with a W or Z boson in pp collisions at13 TeV", CMS-PAS-TOP-17-005

\bibitem{H_4L} CMS Collaboration, ``Measurement of properties of the Higgs boson decaying into the four-lepton final state in pp collisions at 13 TeV", arXiv:1706.09936, submitted to JHEP.

\bibitem{Higgs_mass_r1}  ATLAS and CMS Collaborations, ``Combined Measurement of the Higgs Boson Mass in pp Collisions at $\sqrt{s} \  =  \ 7 {\rm and} 8 {\rm TeV}$ with the ATLAS and CMS Experiments”, Phys. Rev. Lett. 114 (2015) 191803, doi:10.1103/PhysRevLett.114.191803, arXiv:1503.07589.

\bibitem{Higgs_gg} CMS Collaboration, ``Measurement of differential fiducial cross sections for Higgs boson production in the diphoton decay channel in pp collisions at $\sqrt{s}=13~\mathrm{TeV}$", CMS-PAS-HIG-015

\bibitem{Higgs_tautau}  CMS Collaboration,``Observation of the Higgs boson decay to a pair of $\tau$  leptons", arXiv:1708.00373 ; CMS-HIG-16-043-003

\bibitem{Higgs_tt} CMS Collaboration, ``Search for Higgs boson production in association with top quarks in multilepton final states at $\sqrt{s} \ = \ 13 {\rm TeV}$", CMS-PAS-HiG-17-004

\bibitem{WW_scat} CMS Collaboration, ``Observation of electroweak production of same-sign W boson pairs in the two jet and two same-sign lepton final state in proton-proton collisions at 13 TeV", CMS-PAS-SMP-17-004


\bibitem{Higgs_rare_16} CMS Collaboration, `` Search for lepton flavour violating decays of the Higgs boson to $\mu\tau$ and $e\tau$ in proton-proton collisions at $\sqrt{s}$=13 TeV", CMS-PAS-HIG-17-001



\bibitem{Higgs_rare_15}  CMS Collaboration, ``Search for lepton-flavour-violating decays of the Higgs boson”, Phys. Lett. B 749 (2015) 337, doi:10.1016/j.physletb.2015.07.053, arXiv:1502.07400.

\bibitem{P5_prime} CMS Collaboration,``Measurement of the P$_{1}$ and P$_{5}^{\prime}$ angular parameters of the decay $B^{0}\rightarrow K^{*0}\mu^{+}\mu^{-}$ in proton-proton collisions at $\sqrt{s} \ = \ 8 \ {\rm TeV}$, CMS-PAS-BPH-15-005.

\bibitem{DM_search} CMS Collaboration, ``Dark Matter Summary Plots from CMS for LHCP and EPS 2017", twiki.cern.ch/twiki/bin/view/CMSPublic/PhysicsResultsEXO, P 3

\bibitem{SUSY_gen}  CMS Collaboration, ``twiki.cern.ch/twiki/bin/view/CMSPublic/PhysicsResultsSUS\#Run\_2\_Summaries\_13\_TeV, "
plots under section ``Moriond 2017 (36 fb$^{-1}$)"

\bibitem{SUSY_stop} CMS Collaboration, ``Search for supersymmetry using hadronic top quark tagging in 13 TeV collisions", CMS-PAS-SUS-16-050

\bibitem{VLQ_23} CMS Collaboration, ``Search for vector-like quark pair production 
$T\bar{T} (Y\bar{Y}) \rightarrow bWbW$ using kinematic reconstruction in the lepton+jets final state at $\sqrt{s} \ = \ 13  {\rm TeV}$, CMS-PAS-B2G-17-003 

\bibitem{VLQ_53} CMS Collaboration,``Search for top quark partners with charge 5/3 in the single-lepton final state at $\sqrt{s} \ = \ 13 \ {\rm TeV}$",  CMS-B2G-15-006



\bibitem{Long_lived} CMS Collaboration,``Search for stopped long-lived particles produced in pp collisions at $\sqrt{s} \ =  \ 13 {\rm TeV}$", CMS-PAS-EXO-16-004

\bibitem{pixel_tdr} CMS Collaboration, ``CMS Technical Design Report for the Pixel Detector Upgrade", CERN-LHCC-2012-016 ; CMS-TDR-11. - 2012. - 239 p. (Technical Design Report CMS ; 11) 

\bibitem{gem_tdr} CMS Collaboration, ``CMS Technical Design Report for the Muon Endcap GEM Upgrade", CERN-LHCC-2015-012, CMS-TDR-013

\bibitem{hcal_tdr} CMS Collaboration, ``CMS Technical Design Report for the Phase 1 Upgrade of the Hadron Calorimeter", CERN-LHCC-2012-016 ; CMS-TDR-11. - 2012. - 239 p. (Technical Design Report CMS ; 11) 

\bibitem{phase2_tp} CMS Collaboration. ``Technical Proposal for the Phase-II Upgrade of the CMS Detector",  CERN-LHCC-2015-010 ; LHCC-P-008 ; CMS-TDR-15-02. - Geneva : CERN, 2015. - 469 p. (Technical Proposal ; 15.2) 

\bibitem{mtd_ref} CMS Collaboration, ``Initial report of the Fast Timing Working Group", CMS-DP-2016-008 ; CERN-CMS-DP-2016-008







 
\end{thebibliography}
\end{document}